\documentclass[prb,twocolumn,superscriptaddress]{revtex4}

\usepackage{amssymb}
\usepackage{dcolumn}
\usepackage{multirow}
\usepackage{bm}

\usepackage{algorithmic}
\usepackage{tabularx}
\usepackage{setspace}

\usepackage{supertabular}

\usepackage{amsmath,amsfonts}

\usepackage{amsmath}
\usepackage{amssymb}
\usepackage{algorithm2e}
\usepackage{graphicx}
\usepackage{color}
\usepackage{epstopdf}

\newcommand{\f}{\frac}

\begin{document}
\title{Fast and Efficient Stochastic Optimization for Analytic Continuation}

\author{F. Bao}
\affiliation{Computer Science and Mathematics Division, Oak Ridge, Tennessee 37831, USA}
\affiliation{Department of Mathematics, University of Tennessee at Chattanooga
37403, USA}
\author{Y. Tang}
\affiliation{Computer Science and Mathematics Division, Oak Ridge, Tennessee 37831, USA}
\affiliation{Department of Physics, Virginia Tech, Blacksburg, Virginia 24061, USA}
\author{M. Summers}
\affiliation{Computer Science and Mathematics Division, Oak Ridge, Tennessee 37831, USA}
\affiliation{Center for Nanophase Materials Sciences, Oak Ridge National Laboratory,
Oak Ridge, Tennessee 37831, USA}
\author{G. Zhang}
\affiliation{Computer Science and Mathematics Division, Oak Ridge, Tennessee 37831, USA}
\author{C. Webster}
\affiliation{Computer Science and Mathematics Division, Oak Ridge, Tennessee 37831, USA}
\author{V. Scarola}
\affiliation{Department of Physics, Virginia Tech, Blacksburg, Virginia 24061, USA}
\author{T.A. Maier}
\affiliation{Computer Science and Mathematics Division, Oak Ridge, Tennessee 37831, USA}
\affiliation{Center for Nanophase Materials Sciences, Oak Ridge National Laboratory,
Oak Ridge, Tennessee 37831, USA}

\begin{abstract}
The analytic continuation of imaginary-time quantum Monte Carlo data to
extract real-frequency spectra remains a key problem in connecting theory with
experiment. Here we present a fast and efficient stochastic optimization
method (FESOM) as a more accessible variant of the stochastic optimization
method introduced by Mishchenko {\it et al.} \cite{Mishchenko00} and benchmark
the resulting spectra with those obtained by the standard Maximum Entropy
method for three representative test cases, including data taken from studies
of the two-dimensional Hubbard model. We generally find that our FESOM
approach gives spectra similar to the Maximum Entropy results. In particular,
while the Maximum Entropy method gives superior results when the quality of
the data is strong, we find that FESOM is able to resolve fine structure with
more detail when the quality of the data is poor. In addition, because of its
stochastic nature, the method provides detailed information on the frequency
dependent uncertainty of the resulting spectra, while the Maximum Entropy
method does so only for the spectral weight integrated over a finite frequency
region. We therefore believe that this variant of the stochastic optimization
approach provides a viable alternative to the routinely used Maximum Entropy
method especially for data with poor quality.
\end{abstract}

\maketitle

\section{INTRODUCTION}

Quantum Monte Carlo (QMC) methods provide numerically exact results for
interacting quantum many-particle systems and thus are widely used to study
their physics. An important drawback, however, is their inability to directly
give real frequency results, a key limitation considering the large number of
experiments that measure dynamic quantities. From the imaginary time QMC data,
the real frequency spectrum $A(\omega)$ has to be recovered through the process
of analytic continuation, a highly ill-posed inverse problem that remains a key
stumbling block in connecting theory with experiment.

To address this challenge it has proven useful to employ a framework based on
Bayesian statistical inference. The state-of-the-art and most widely used tool
based on Bayesian statistics is the Maximum Entropy (MaxEnt) method
\cite{Skilling84} pioneered by Silver, Sivia, Jarrell and Gubernatis
\cite{Silver90a,Silver90b,Gubernatis91,Jarrell96} for applications in this
area. It introduces an entropy-like regularization term that measures the
deviation from a default spectrum, and then obtains the most probable spectrum
through a deterministic optimization process. Another method that uses explicit
bur adjustable regularization through the use of consistent constraints was
recently introduced by Prokof'ev and Svistunov \cite{Prokofev13}.

An alternative stochastic method was developed by Sandvik \cite{Sandvik98}, in
which a fictitious temperature is introduced to define the probability of a
given spectrum by a Boltzmann weight. This allows for efficient Monte Carlo
sampling of possible spectra from which the final spectrum is obtained as a
weighted average. A refined version of this approach, which, similar to MaxEnt,
uses a default model, was later introduced by Beach \cite{Beach04} and shown to
become formally equivalent to the MaxEnt method if the fictitious system is
treated at a mean-field level. In addition, Fuchs et al.\cite{Fuchs10} showed
that the fictitious temperature introduced in this algorithm can be eliminated
based on principles of Bayesian statistical inference in a similar fashion as
the regularization parameter of the MaxEnt approach is removed.

Mishchenko et al. \cite{Mishchenko00} used a similar idea to set up a
stochastic optimization method (SOM) that randomly samples solutions with a
certain weight but without interpretation of the weights as a Boltzmann
distribution. In this approach, one randomly samples a large enough number of
possible solutions $A(\omega)$, each of which optimizes the deviation from the
QMC data, but allows for solutions with larger deviation to implicitly
regularize the problem. One important feature of this approach is that it uses
a different and much more complex parametrization of the spectrum that does not
impose a rigid, discrete frequency grid and allows for overlapping rectangles
from which the spectrum is composed. While this allows for more flexibility in
the solution, it leads to a complex update algorithm and a very large search
space that is difficult to manage. 

Here, we introduce a fast and efficient stochastic optimization method (FESOM)
as an accessible variant of Mishchenko's original SOM that is based on the
same idea, i.e. a stochastic sampling of possible spectra. But instead of the
complex parametrization introduced by Mishchenko et al., it uses the usual
parametrization of solutions $A(\omega)$ in terms of a discrete frequency
grid, resulting in a more manageable algorithm. We apply this approach to a
number of representative problems and compare the results against those
obtained from standard MaxEnt calculations. We include two test cases of
approximate spectral functions derived from the two-dimensional Hubbard model
on a square lattice. In the following section, we review the state-of-the art
MaxEnt method that we use to benchmark our approach. We then discuss the new
FESOM in Sec.~\ref{SOM}, and show the results of three different numerical
examples in Sec.~\ref{cases}.

\section{Analytic continuation and Bayesian statistics}

The analytic continuation process involves an inversion of the integral
\begin{equation}\label{eq:fredholm}
	G(i{\omega}_n) = \int d\omega\, K(i\omega_n, \omega) A(\omega)\,.
\end{equation}
Here, $G(i\omega_n)$ is an observable such as the single-particle Green' s
function measured in a QMC calculation as a function of discrete Matsubara
frequencies $\omega_n$ on the imaginary axis, $A(\omega)=-1/\pi\, {\rm Im}\,
G(\omega)$ is the spectral function and quantity of interest, and
$K(i\omega_n,\omega)$ is the kernel. For fermionic Green's function considered
here one has $\omega_n = (2n+1)\pi T$ for a temperature $T$, and the kernel
takes the form
\begin{equation}
\label{eq:kernel}
	K(i\omega_n, \omega) = \frac{1}{i\omega_n -\omega} \,.
\end{equation}
After discretization of the real frequency axis into $L$ intervals,
$\{\omega_l\}_{l=0}^{L}$, Eq.~(\ref {eq:fredholm}) is written in matrix-vector
form
\begin{equation}\label{eq:lineareq}
G_n = \sum_{l=1}^L K_{nl}A_l
\end{equation}
with $K_{nl} \equiv \Delta\omega_l/(i\omega_n-\omega_l)$, $G_n \equiv G
(i\omega_n)$ and $A_l\equiv A(\omega_l)$ and the frequency intervals
$\Delta\omega_l=\omega_{l+1}-\omega_l$. The difficulty of inverting
Eq.~(\ref{eq:fredholm}) arises from the small tails in the kernel function at
large frequencies $\omega$. In other words, the matrix $K_{nl}$ is
ill-conditioned, i.e. small changes or statistical errors in the QMC data $G_n$
cause large errors in the quantity of interest $A_l$ and there are an infinite
number of solutions.

Approaches that address this problem can be formulated in terms of Bayesian
statistical inference, in which one considers the Bayesian formula
\begin{equation}\label{eq:Bayesian}
P(A | G) \propto P(G|A) P(A)\,.
\end{equation}
Here, $P(A|G)$ is the posterior probability of the spectrum $A$ given the data
$G$, the prior probability $P(A)$ encodes prior information about $A$ and the
likelihood function $P(G|A)$ measures the quality of the fit between $G$ and
$KA$. The problem of finding the most probable spectrum $A$ given the data $G$
is thereby converted into the much easier problem of optimizing the likelihood
function and prior probability. One can then select the most probable spectrum
$A$ that maximizes $P(A|G)$ as in the case of the MaxEnt method, or obtain the
final spectrum $A$ from a weighted average over possible solutions
\begin{equation}
	\label{eq:averaA}
	\bar{A} = \int dA\, A\, p(A|G)
\end{equation}
as in the case of the stochastic methods.

% Typically, QMC method gives several samples of Green's function data and we
% denote $G_m^{k} := G^{k}(i\omega_m)$ the $k$-th sample of the QMC data for $m
% = 1, 2, \cdots, M$ and $K_{mn}$ is the value of kernel $K$ at
% $i\omega=i\omega_m$ and $\omega = \omega_n$. The goal of the analytic
% continuation problem is to find the spectral function $A$ from the given QMC
% data $G := \{G^{k}\}_{k=1}^{K}$, where $G^{k} : = \{ G_1^k, G_2^k, \cdots,
% G_M^k\}$ and $K$ is the total number of QMC samples. The major difficulties
% for this type of problems is the kernel $K$ is an ill-conditioned matrix and
% the QMC data is typically perturbed by noise. Therefore, solving the spectral
% function $A$ by multiplying the inverse kernel $K^{-1}$ directly to data $G$
% would get a noisy and non-unique solution from different realizations of data
% which makes the analytic continuation problem ill-posed.

\section{Maximum Entropy}
\label{maxent}

The MaxEnt approach \cite{Jarrell96} uses the Bayesian statistical inference
formula, Eq.~(\ref{eq:Bayesian}) to find the most probable spectrum $A$ given
the input data $G$. This is done by maximizing both the likelihood function
$P(G|A)$ and the prior probability $P(A)$.

The likelihood function $P(G|A)$ is defined according to the central limit
theorem as
\begin{equation}
	\label{eq:likelihood}
	P(G|A) = e^{-\chi^2/2}\,,
\end{equation}
where
\begin{equation}\label{eq:chi2}
\chi^2[A] = \frac{1}{N}\sum_{n=1}^N\left(\frac{G_n-\sum_l K_{nl}A_l}
{\sigma_n}\right)^2
\end{equation}
encodes the quality of the fit of the data $G$ by the spectrum $A$. Here
$G_n = 1/N_s\sum_{i=1}^{N_s}G^i_n$ is obtained as the mean value of a
number $N_s$ of different QMC samples with $G_n^i\equiv G^i(i\omega_n)$ the
$i$-th sample, and the variance
\begin{equation}
	\label{covariance}
	\sigma_n^2 = \frac{1}{N_s-1}\sum_{i=1}^{N_s}(G_n^i-G_n)^2\,.
\end{equation}
Note that this form assumes that no correlations between different frequencies
$i\omega_n$ are present in the QMC data $G_n$. When there are correlations, the
covariance matrix has to be diagonalized and both the data $G_n$ and the kernel
$K$ have to be rotated into this diagonal representation \cite{jarrellBook}.

A simple minimization of $\chi^2$ with a least-square fit of the data ${G}$
with $KA$ leads to noisy and an infinite number of non-unique solutions.  The
MaxEnt method addresses this problem by regularization of the least-square fit.
It introduces a prior distribution
\begin{equation}
	\label{eq:prior}
	p(A) = e^{\alpha S[A]}\,,
\end{equation}
with $\alpha$ a positive constant, the regularization parameter, and
\begin{align}\label{eq:entropy}
S[A] &= -\int d\omega \left[A(\omega)- D(\omega) - A(\omega)\ln\frac{A(\omega)}
{D(\omega)}\right]\nonumber\\
&= -\sum_{l=1}^L \left[A(\omega_l)-D(\omega_l)-A(\omega_l)\ln \frac{A(\omega_l)}
{D(\omega_l)}\right]\Delta\omega_l
\end{align}
an entropy like term defined relative to a positive definite and normalized
function $D(\omega)$, the default model. Thus, in order to maximize the
posterior probability $p(A|G)$, the MaxEnt minimizes the function
\begin{equation}\label{functional}
	Q[A] = \frac{1}{2}\chi^2[A] - \alpha S[A]\,.
\end{equation}
The Bayesian inverse optimization of the posterior probability $p(A|G) \propto
e^{-Q[A]}$ hence becomes a deterministic optimization for the regularized form
$\f{1}{2}\chi^2[A] - \alpha S[A]$ as a standard optimization problem. Here
$\alpha$ mediates the competition between the $\chi^2$ fit of the data and
prior information contained in $S[A]$. It is the $\chi^2[A]$ term that ensures
that the spectral function will give a good fit of the data, while the $S[A]$
term avoids over-fitting of the data by guiding $A(\omega)$ towards a default
model given by $D(\omega)$. The Bayesian inference formulation also allows to
eliminate the free parameter $\alpha$ by calculating the posterior probability
of $\alpha$, i.e. $P(\alpha|G)$.  One can then perform the MaxEnt
procedure for different values of $\alpha$ to give estimates for the spectrum
$A_\alpha$ and then select the most probable $A_\alpha$ that corresponds to the
maximum $p(\alpha|{\bar G})$. Here we use Bryan's method \cite{Bryan90}, in
which one averages over all spectra $A_\alpha$ weighted by the posterior
probability of $\alpha$ to obtain
\begin{equation}
	A = \int d\alpha\, P(\alpha|G) A_\alpha\,.
\end{equation}

\section{Fast and Efficient Stochastic Optimization Method}
\label{SOM}

An alternative numerical approach to solve the analytic continuation problem
are the stochastic inference method introduced by Sandvik \cite{Sandvik98} and
refined by Beach \cite{Beach04} and Fuchs et al. \cite{Fuchs10} and the
stochastic optimization method developed by Mishchenko \cite{Mishchenko00}.
While these approaches can outperform the traditional MaxEnt method and yield
spectra with more features and less regularization, they can be very
numerically expensive. Fuchs et al. \cite{Fuchs10} commented that the
necessity to perform calculations for a wide range of regularization
parameters in their refined approach can lead to run-times of 20 processor
hours. Similarly, the complexity of the parametrization of the spectrum used
in Mishchenko's SOM and the associated extensive search space seems comparably
expensive. Here we discuss an efficient and more accessible variant of this
stochastic optimization method that also uses a Bayesian framework for the
analytic continuation problem with only minimal prior information on the
spectrum.

In many situations, one has only minimal pre-knowledge of the prior probability
of $A$, i.e. $p(A)$. Therefore, we assume that the prior distribution $p(A)$ is
uniform and the posterior distribution is equivalent to the likelihood, i.e.
$p(A | G) = p(G | A)$.  The most straightforward way to construct an empirical
distribution is to use the Markov Chain Monte Carlo (MCMC) sampling method
\cite{mcmc}. However, the dimension of the distribution is the partition number
$L$ of the frequency $\omega$ which is typically large. In this case, the MCMC
sampling method becomes very inefficient, especially in simulating the
statistically insignificant region due to the low acceptance rate. To overcome
this problem in MCMC type methods, we propose an efficient scalable numerical
algorithm which constructs an empirical distribution for $p(G | A)$  with
independent random samples.

The central idea of our algorithm is to build the target probability
distribution $p(A|G)$ by running several parallel optimization procedures. To
this end, we compute $J$ realizations of optimal spectral functions $A$ based
on QMC data $G$ in a stochastic manner and use the distribution of all $J$
realizations of stochastically optimized spectral functions to be a
representation of the distribution for $A$. For each realization, the
stochastic optimization aims to minimize the $\chi^2$ error and the random
optimal spectral function will be very noisy due to the fact that the analytic
continuation problem is ill-posed. Also, different realizations have very
different features. However, since all random samples of the spectral function
have very small $\chi^2$ error, statistically they capture the feature of the
true spectral function $A$ and the mean value of all the samples will be a good
estimate for the final spectral function $A$.

Specifically, for any given initial guess of the spectral function $A(\omega)$,
which we denote by $D(\omega)$, we introduce an initial partition $\Pi_0$ of
the frequency axis defined by
$$
\begin{aligned}
\Pi_0 := \{ \omega_l & | a = \omega_0 \leq \omega_1 \leq \omega_2\\
& \ \  \leq \cdots \leq \omega_l \leq \omega_{L-1} \leq \omega_L = b \  \}.
\end{aligned}
$$
Here $a$ is the
lower boundary of the test frequency region and $b$ the upper boundary. In many
cases, one is more interested in resolving features in the low frequency
region. Therefore, we let the partition stepsize grow exponentially with
increasing absolute value of the frequency \footnote{For example, we chose the
stepsize $0.1$ for the interval $[-2, 2]$, $0.2$ for $[-4, -2]$ and $[2, 4]$,
$0.4$ for $[-6, -4]$ and $[4, 6]$, and so on.}. Thus the frequency stepsize
$\Delta \omega_l$ is small for small $|\omega|$ and grows with $|\omega|$, so
that we have better resolution in the more important low frequency region.

With the initial guess $D(\omega_l)$ and the partition $\Pi_0$,  we initialize
$R$ realizations of spectral functions $A$, denoted by $\tilde{A}^r_0$, $r = 1,
\cdots, R$, with 
$$\tilde{A}^r_0(\omega_l) = D(\omega_l),  \quad l = 0, 1,
\cdots, L. 
$$ 
Each realization is initialized with the same $D(\omega_l)$, for which we
typically choose a Gaussian in the absence of external information. For each
realization starting with $\tilde{A}^r_0$, we run an independent stochastic
optimization procedure to minimize the $\chi^2$ error and update the simulated
spectral function $\tilde{A}^{r}_i$ from iteration step $i$ to $i+1$, where $i
= 0, 1, 2, \cdots$. We find that if we run enough iterations steps the final
result does not depend on the initial function $D$.

Suppose we have the $r$-th realization of the simulated spectral function at
iteration step $i$, i.e. $\tilde{A}^{r}_i$. To find an optimal solution, we add
a Gaussian process, denoted by the random vector of length $L$, $\lambda^{r}_i :=
(\lambda^{r}_i(\omega_1),\lambda^{r}_i(\omega_2), \cdots,
\lambda^{r}_i(\omega_L))$, to $\tilde{A}^{r}_i$ and get a proposed spectral
function
\begin{align}
\tilde{A}^{r}_{i+\f{1}{2}} &= \f{1}{I}\left( \tilde{A}^{r}_i + \lambda^
{r}_i\right)\,.
\end{align}
Here the constant $I$ is chosen so that the spectrum $\tilde{A}^r_
{i+\f{1}{2}}$ is normalized, i.e. satisfies
\begin{align}
	\sum_{l=1}^L \tilde{A}^r_{i+\frac{1}{2}}(\omega_l)\Delta\omega_l = 1\,.
\end{align}

In principle, the only constraint we impose on the random variables
$\lambda^r_i$ is that the proposal spectrum is positive definite, i.e.
$\tilde{A}^r_{i+\f{1} {2}} \geq 0$. However, in order to allow for implicit
regularization and to improve efficiency, we normally set the Gaussian process
$\lambda^{r}_i$ to a multi-variate Gaussian random variable with mean zero
and covariance $C$, which determines the smoothness of the noise $\lambda^r_i$
as a function of frequency $\omega_l$. If the correlation is strong (large
$C$), the noise we add is smooth; for small $C$ the noise fluctuates strongly
between neighboring frequencies. Thus, the covariance $C$ may be considered a
smoothing factor, which provides an implicit regularization. Since the
partition step-size on the frequency axis restricts the resolution of possible
features in the spectral function, we let the covariance function $C$ depend
on the partition of the frequency. There are many choices of the covariance
function, including constant, linear, squared exponential, Ornstein-Uhlenbeck,
rational quadratic or other forms. Here, we choose an Ornstein-Uhlenbeck form,
i.e.%\begin{equation}\label{covariance}
%K(\omega_{n_1}, \omega_{n_2}) = \f{1}{1 + | n_1 - n_2 | \alpha},
%\end{equation}
\begin{equation}\label{covariance}
C(\omega_{l_1}, \omega_{l_2}) = \exp( - \alpha| l_1 - l_2 | ),
\end{equation}
where $\alpha$ is a positive constant. A popular choice of $\alpha$ is
provided by the ``maximum posteriori estimate'', which is a mode of the
posterior distribution  \cite{Harold80}. Note that we let the noise
correlations depend on the number of intermediate partition steps, $|l_1 -
l_2|$, instead of the frequency directly. This means that the effective
correlation between frequencies changes with frequency since the resolution of
our frequency grid changes. It is small in the low frequency region where the
step size is small and the resolution is high, while the effective correlation
is high in the larger frequency region where the step size is large. This
frequency adaptive noise correlation is consistent with the idea that finer
structures are to be resolved in $A (\omega)$ in the more important low
frequency region, while stronger smoothening can take place in the higher
frequency region. In general, larger values of $C$ will impose more smoothing
on individual realizations $\tilde{A}^r(\omega_l)$ and thus reduce the number
of realizations needed to obtain a smooth average $\bar{A}(\omega)$. Thus, the
correlation parameter $C$ may be used as a tuning parameter to balance the
gain in details in $A(\omega)$ against an increase in computer time. Given the
covariance matrix $C(\omega_{\ell_1},\omega_{\ell_2})$ in
Eq.~\eqref{covariance}, we then generate the random vector $\lambda^r_i$ from
the $L$-dimensional joint normal distribution $N(0,C)$ with mean $0$ and
covariance $C$.

% With the proposed spectral function ${\tilde A}^j_{i+\f{1}{2}}$, we then compute
% the simulated data $\tilde{G}_{i+\f{1}{2}}^{j}(\omega_m) = \sum_n K_{mn} {\tilde
% A}^j_{i+\f{1}{2}}(\omega_n)\Delta \omega_n$ and compare its $\chi^2$ error
% $$
% \chi^2[\tilde{A}^{j}_{i}] : = (\tilde{G}^{j}_i  - \bar{G} ) ( 2 \Sigma )^{-1} (\tilde{G}^{j}_i - \bar{G} )^{T}
% $$
% with that of the previous spectrum ${\tilde A}^j_{i}$. Here, $\bar{G}$ is the
% mean value of the QMC samples for $G$.

% If the $\chi^2$ error corresponding to the proposed spectral function
% $\tilde{A}^{j}_{i+\f{1}{2}}$ is smaller then the $\chi^2$ error corresponding
% to the prior spectral function $\tilde{A}^{j}_{i}$, we consider the proposed
% spectral function $\tilde{A}^{j}_{i+\f{1}{2}}$ to be closer to the true spectral
% function $A$ and accept $\tilde{A}^{j}_{i+\f{1}{2}}$ as our simulated spectral
% function in iteration step $i+1$, i.e. $\tilde{A}^{j}_{i+1} : =
% \tilde{A}^{j}_{i+\f{1}{2}}$; otherwise, we consider the proposed spectral
% function $\tilde{A}^{j}_{i+\f{1}{2}}$ is not as good as the prior choice
% $\tilde{A}^{j}_{i}$ and we keep the prior spectral function as our simulated
% spectral function in iteration step $i+1$, i.e. $\tilde{A}^{j}_{i+1} : =
% \tilde{A}^{j}_{i}$.

If the proposed spectral function $\tilde{A}^r_{i+\frac{1}{2}}$ fits the data
better than the previous $\tilde{A}^r_{i}$, i.e. if $\chi^2[\tilde{A}^r_
{i+\frac{1}{2}}] < \chi^2[\tilde{A}^r_{i}]$, we accept the update and set
$\tilde{A}^r_ {i+1} = \tilde{A}^j_{i+\frac{1}{2}}$. Otherwise, the update is
rejected and $\tilde{A}^r_{i+1} = \tilde{A}^r_{i}$. Thus, the $\chi^2$ error
between the simulated Green's function $\tilde{G}$ and the QMC experimental
data $G$ will decrease monotonically. In our implementation, the
optimization process is stopped in the $j$-th iteration if $\chi^2(\tilde{A}^
{r}_j) \leq \epsilon$ for a fixed threshold $\epsilon$. $\tilde{A}^{r} : =
\tilde{A}^{r}_j$ then denotes the final spectral function for realization $r$.
From a number $R$ of independent stochastic optimization procedures, we obtain a
set of random optimal spectral functions, i.e. $\{\tilde{A}^r\}_{r=1}^R$,
which forms an empirical distribution for the spectral function $A$, denoted
by $P_0(\tilde{A} | G)$. We note that the stochastic optimizations for
different realizations are independent, which makes the algorithm scalable in
the stochastic optimization procedure.

The threshold $\epsilon$ is a user defined positive constant, which should be
chosen according to the complexity of the problem. In practice, we keep
$\epsilon$ of the same order as the variance of the QMC data, in order to
avoid overfitting of the data and to keep the efficiency of the optimization
process high. In contrast to the SOM used by Mishchenko {\it et al.}
\cite{Mishchenko00}, we do not allow for updates that increase $\chi^2$. The
fact that we use a random, global process to update the spectral function,
however, provides a means to get out of local minima with high $\chi^2$. In
spite of this, the optimization process slows down at very small $\chi^2$,
because the probablity of finding a better solution is small. As we observe
in practice, keeping $\epsilon$ of similar size as the QMC errors ensures that
the process does not become prohibitively inefficient.

% \begin{rem}
% It's worth to point out that the proposed spectral function
% $\tilde{A}^{j}_{i+\f{1}{2}}$ is a random process and it always has a chance to
% reduce the $\chi^2$ error, thus would give opportunities to get out of local
% minima. However, as a stochastic optimization method, when the $\chi^2$ is
% getting smaller and smaller, the chance we can get a
% $\tilde{A}^{j}_{i+\f{1}{2}}$ that would give a smaller $\chi^2$ is also
% smaller and smaller.
% \end{rem}
% \begin{rem}
% The threshold $\epsilon$ is a user defined positive constant which depends on the complexity of the problem. If we expect there are many complex features in the spectral function, typically we choose a smaller $\epsilon$.
% \end{rem}

The random optimal spectral function that results from a single realization
may not capture all the important features in the true spectral function and
will be noisy due to the fact that the problem is ill-posed. The weighted
average, Eq.~\eqref{eq:averaA}, of the different realizations, however, will
be smooth if the number of realizations is large enough. Since we stop each
optimization when $\chi^2$ reaches the same value $\epsilon$, the weights in
Eq.~\eqref{eq:averaA} are all identical and the final spectrum $\bar
{A}(\omega_n)$ is obtained from a simple average
\begin{align}
\label{eq:averageA}
\bar{A}(\omega_l) =
	\f{1}{R}\sum_{r=1}^R \tilde{A}^r(\omega_l)\,.
\end{align}
Note that in practice, we chose a maximum number $S$ of update steps. If a
particular optimization procedure for realization $r$ does not reach $\chi^2
\leq \epsilon$ in $S$ steps, the update process is stopped and the spectrum in
step $S$, $\tilde {A}^r_S$ is used as the final result $\tilde{A}^r$ for this
realization. In this case, we still use Eq.~\eqref{eq:averageA} to compute the
final spectrum $\bar{A}(\omega_l)$ and assume that it is accurate enough. For
the examples we considered in Sec.~\ref{cases}, however, we find that $\chi^2$
always reaches $\epsilon$ before $S$ updates, so that this is not an issue. We
generally choose the number $R$ of realizations large enough to get a smooth
final solution for the spectrum. $R$ is generally problem specific and also
depends on the parameter $\alpha$ in the correlation between neighboring
frequencies as will be discussed in Sec.~V Example 1.

It is important to point out that the original frequency partition $\Pi_0$ is
not informed by the data $G$ and thus is not adaptive to the features in the
spectral function. Because of the stochastic nature of the SOM procedure, one
has a representation for the data informed distribution $\tilde{P}_0(A | G)$ in
addition to the approximate spectral function $\bar{A}$. From the distribution,
one can get the standard deviation for every single frequency in $\Pi_0$ and
from that construct a confidence band for the estimate spectral function
$\bar{A}$. A wide confidence band indicates large fluctuations in the different
realizations, which may point to possible fine structure in the true spectral
function $A$. Based on the width of the confidence band, one can then modify
the frequency partitioning to allow the algorithm to resolve more detail in the
solution. If the confidence band is wide in a certain frequency region, we use
more partition points in that region, and converseley, if the confidence band
is narrow, we use less partition points. We then re-run the stochastic
optimization procedure with the modified frequency partition. The complete
algorithm for fixed frequency partitioning is summarized in Algorithm 1.

\begin{table}
\begin{supertabular}{p{0.4\textwidth}}
\hline\noalign{\smallskip}
{\bf Algorithm 1}: {\em Stochastic optimization method}\\
\noalign
{\smallskip}\hline
\noalign{\smallskip}
\begin{spacing}{1.1}
\begin{algorithmic}\label{algorithm}
\item[\textbf{Initialize}] Choose partition $\Pi_0$ for frequency
grid $\omega_l$, initial guess of the spectral function $D(\omega_l)$, sample
size $R$, $\chi^2$ error threshold $\epsilon$ and optimization update step
number $S$.
% \item[\textbf{while}] $t =1, 2, \cdots, T$, \textbf{do} \\
% \vspace{-0.3cm}
\begin{description}
\item[\textbf{for}] $r=1, 2, \cdots, R$ \\
 Let $\tilde{A}^r_0 = D$ on partition $\Pi_0$ \\
\vspace{-0.3cm}
\begin{description}
	\item[\textbf{while}]  $1 \leq i < S $ \textbf{do}  \\
$i=i+1$\\
Propose $\tilde{A}^{r}_{i+\f{1}{2}} =\tilde{A}^{r}_{i} + \lambda^r_i$ \\
Compute $\chi^2(\tilde{A}^{r}_{i})$ and $\chi^2(\tilde{A}^{r}_{i+\f{1}{2}})$\\
\textbf{if} $\chi^2(\tilde{A}^{r}_{i+\f{1}{2}}) \geq \chi^2(\tilde{A}^{r}_{i}) $ \\
\quad $ \tilde{A}^{r}_{i+1} = \tilde{A}^{r}_{i}$ \\
\textbf{else}\\
\quad $\tilde{A}^{r}_{i+1} = \tilde{A}^{r}_{i+\f{1}{2}}$ \\
\textbf{end if}\\
\textbf{if} $\chi^2(\tilde{A}^{r}_{i+\f{1}{2}}) \leq \epsilon$ \\
\quad $\tilde{A}^r = \tilde{A}^{r}_{i+1}$ \\
\quad Break \\
\textbf{end if}\\
\vspace{-0.1cm}
\item[\textbf{end while}]
\vspace{-0.1cm}
\end{description}
%\vspace{-0.35cm}
%Let $x_{n+\f{1}{2}}^i = \tilde{x}_M^{i}$
\vspace{-0.4cm}
\item{\textbf{end for}}
\end{description}
\vspace{-0.2cm}
Approximate the empirical distribution $P(\tilde{A})$ \\ % with samples $\{\tilde{A}^{j}\}_{j=1}^{J}$ \\
% Update $\Pi_t$ based on $P_t(\tilde{A})$
% \item[\textbf{end while}]
\end{algorithmic}
Compute $\bar{A}$ from $P(\tilde{A})$ according to Eq.~\eqref{eq:averageA}.

\vspace{-0.4cm}
\end{spacing}
\\
\hline
\end{supertabular}
\end{table}

\section{Numerical Examples}
\label{cases}

%{\color{red} Some comments (more to follow as we update the draft): \\
%\begin{itemize}
%\item We don't need a separate Fig.1, and Fig.2, 3 and 4 can be combined into
%one figure with 3 panels. We should also show the exact spectrum in the last
%panel (what is now Fig. 4))
%\item I think we should switch Examples 2 and 3, so that the first 2 examples
%are synthetic, and the last example is from actual QMC data
%\item For the synthetic examples, we have to state how the different samples for
%the input data $G(i\omega_m)$ were generated.
%\item It would be nice if we could state the actual $\chi^2$ errors for MaxEnt
%and SOM. Ideally, the SOM should have smaller $\chi^2$ since it does not use
%explicit regularization.
%\end{itemize}}
In this section, we discuss the results of applying our FESOM variant to three
different numerical examples to assess its effectiveness and compare with
results obtained from the standard MaxEnt procedure. The first two examples
are cases for which the exact spectrum $A(\omega)$ is known and different
samples for the data $G (i\omega_n)$ are generated synthetically. The third
case is a problem for which $G (i\omega_n)$ is generated from a QMC simulation
of a single-band Hubbard model and the true spectrum $A(\omega)$ is unknown.
For all three examples in this paper, we choose the FESOM regularization
parameter for the noise correlations $\alpha = 0.5$.

\subsection*{Example 1.}
\begin{figure}[h!]
{\includegraphics[scale = 0.45]{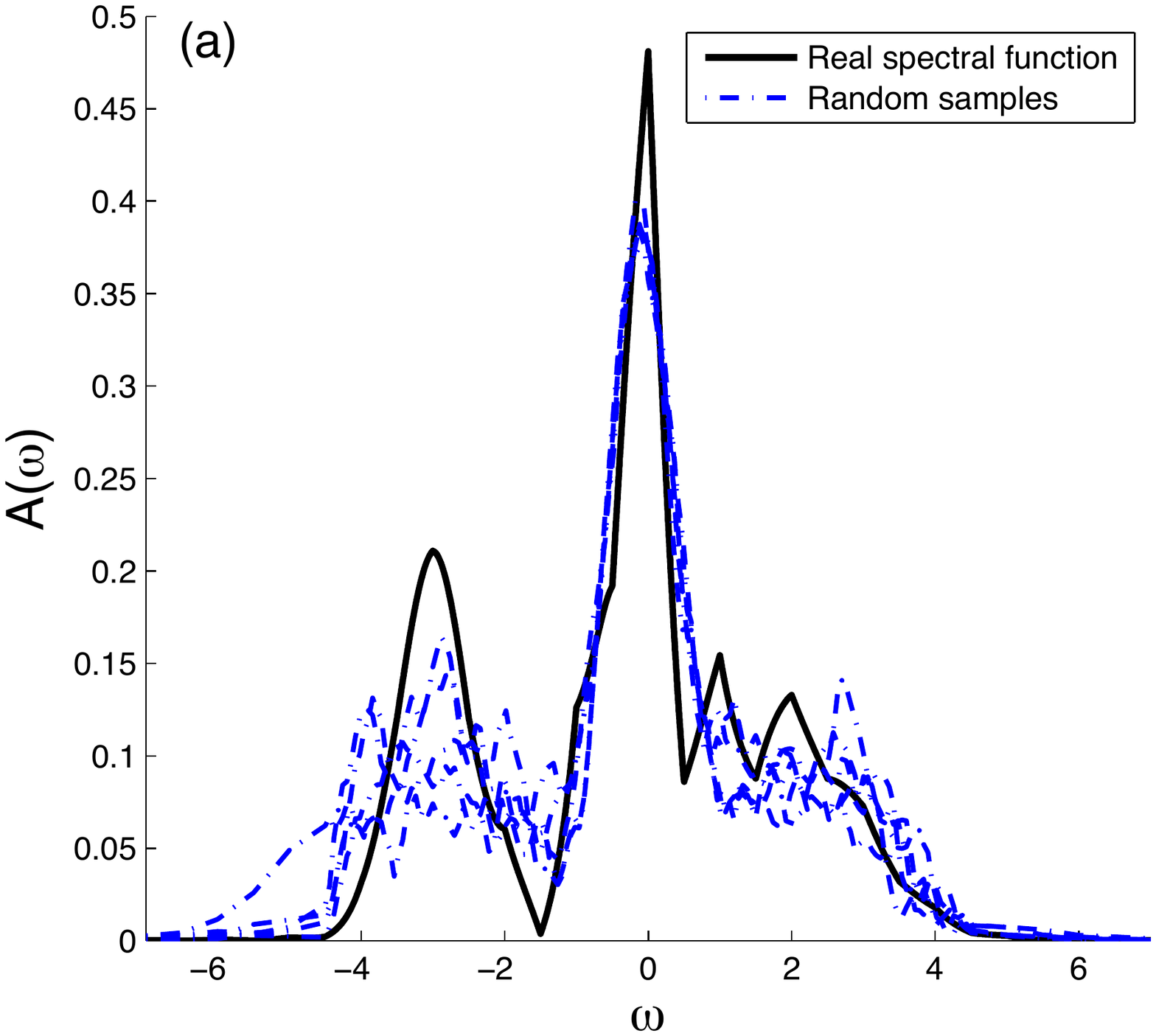}}
{\includegraphics[scale = 0.45]{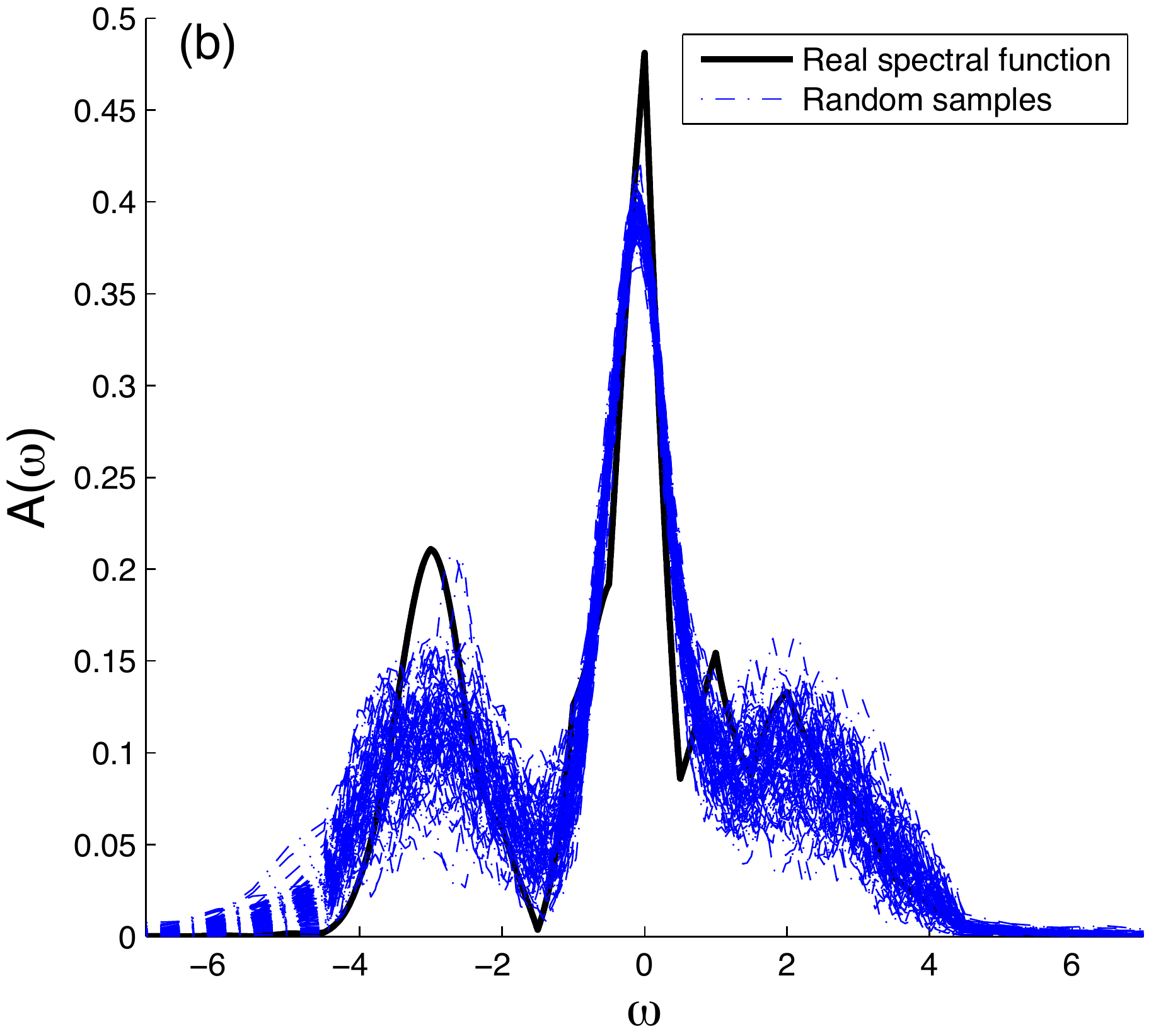}}

\caption{Example 1. The synthetic spectral function $A(\omega)$ (black line) is
used to generate different samples for the input data $G(i\omega_n)$ using
Eq.~\eqref{eq:fredholm} and compared to the results of $5$ (a) and $100$
(b) independent realizations of stochastic optimal spectral function
samples (blue dashes curves). Here we have used a
$\chi^2$ threshold $\epsilon=0.05$.}
\label{fig:Ex1_Samples}
\end{figure}

In this example, we consider a synthetic problem for which we make up a
spectrum $A(\omega)$ with features similar to those expected for the electronic
spectral function of a metallic system with a pronounced quasiparticle peak at
the Fermi energy $\omega=0$ (black line in Fig.~\ref{fig:Ex1_Samples}). From
this spectrum $A(\omega)$ we generate the input data $G(i\omega_n)$ using the
Hilbert transform in Eq.~\eqref{eq:fredholm} and setting the temperature
$T=0.1$. We then generate 1000 samples of $G(i\omega_n)$ by adding noise, i.e.
the i$^{th}$ sample $G^i_n$ is obtained as $G^i_n = [\sigma_{\rm noise}\, N(0,1) +1]G_n$,
where $N(0,1)$ is noise drawn from a normal distribution with mean 0 and
standard deviation 1 and $\sigma_{\rm noise}$ is the noise amplitude.

% \begin{figure}[h!]
% \begin{center}
% \includegraphics[scale = 0.4]{./Ex1_A}
% \end{center}
% \caption{ Example 1. Synthetic spectral function $A$. }\label{Ex1_A}
% \end{figure}

We first illustrate the behavior of individual realizations of the
stochastically optimized spectral function $\tilde{A}(\omega)$ for input data
with relatively low quality, which we generated using a noise amplitude $\sigma_{\rm
noise}=0.1$. We have set the $\chi^2$ threshold $\epsilon = 0.05$. In the top
panel of Fig.~\ref{fig:Ex1_Samples} we compare the true spectrum $A(\omega)$
(black line) with $5$ different realizations of $\tilde{A}(\omega)$ (blue
dashed curves). As one sees, all realizations capture the central peak in the
true spectral function, but different realizations have different features in
the higher frequency region. The bottom panel of Fig.~\ref{fig:Ex1_Samples}
shows $100$ different realizations of the stochastic optimal spectral function
samples, again compared with the true spectral function $A(\omega)$. Here one
sees again that the center peak is well capture by all samples with very little
difference between the samples. In addition, the plots show that statistically,
the samples capture the peak on the left as well as the fluctuations on the
right.

%\begin{figure}[h!]
%\begin{center}
%\includegraphics[scale = 0.5]{./Ex1_HighSamples}
%\end{center}
%\caption{ Example 1. $100$ realizations of stochastic optimal spectral function samples.  }
%\end{figure}

As discussed in Sec.~\ref{SOM}, we use a Gaussian process, in which the added
noise is correlated between adjacent frequencies, to propose updates to the
spectral function in the stochastic optimization procedure. This leads to
proposals that are significantly smoother than what one would get if the noise
added to different frequency points was uncorrelated. This improves the
efficiency of the algorithm since fewer realizations are needed to get a
smooth average $\bar{A} (\omega)$. This benefit, however, comes at the cost of
losing possible fine structure details, which are potentially flattened out by
the correlated noise.

\begin{figure}[h!]
\begin{center}
\includegraphics[scale = 0.55]{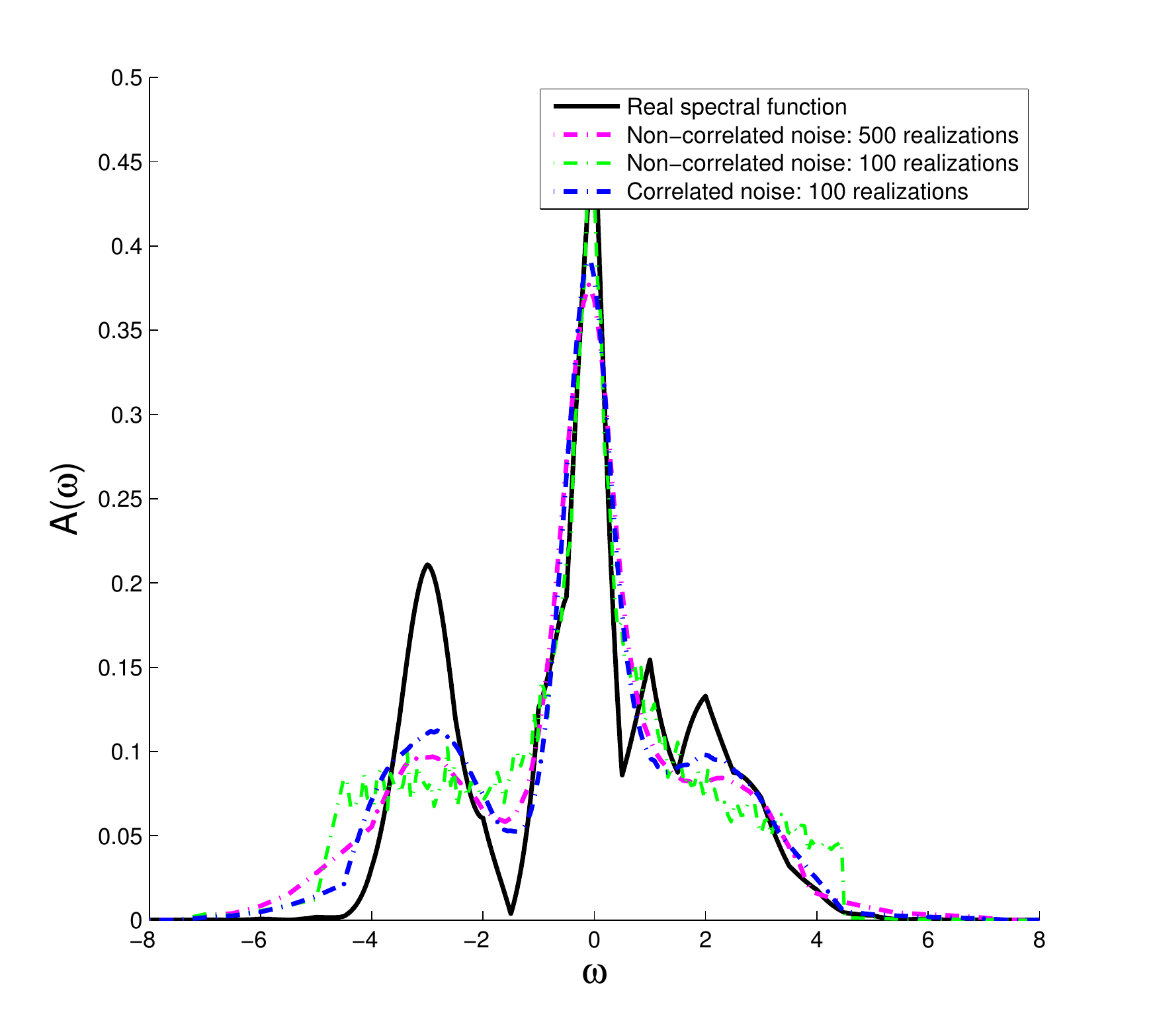}
\end{center}
\caption{Example 1. Averaged spectrum $\bar{A}(\omega)$ obtained from
the stochastic optimization with 500 realizations and non-correlated noise
proposals ($\alpha=0$, magenta dash-dotted line) compared with the result from
100 realizations and non-correlated noise proposal ($\alpha=0$, green
dash-dotted line) and 100 realizations correlated noise proposals
($\alpha=0.5$, blue dash-dotted line). Here we have used a $\chi^2$ threshold
$\epsilon=0.05$. }\label{Ex1_Comparison_Smooth}
\end{figure}

To illustrate the effect of this implicit regularization, we compare in
Fig.~\ref{Ex1_Comparison_Smooth} the simulated spectral function obtained by
using correlated noise proposals (blue dash-dotted line) with that obtained
from using non-correlated noise proposals (magenta dash-dotted line). We have
found that in the case of correlated noise proposals, 100 realizations are
sufficient to give a smooth final average $\bar{A}(\omega)$, while the case of
non-correlated noise proposals required 500 realizations. As one sees from the
green dash-dotted line, 100 realizations are not sufficient to give a smooth
result when the noise proposals are uncorrelated. With correlated noise
proposals, however, 100 realizations provide a smooth result for $\bar
{A}(\omega)$, that is very similar to the result with uncorrelated noise and
500 realizations as well as to the true spectrum $A(\omega)$. It is also clear
that in this case, the correlations in the noise do not result in any loss of
detail in $\bar{A}(\omega)$.

\begin{figure}[h!]
\begin{center}
\includegraphics[scale = 0.6]{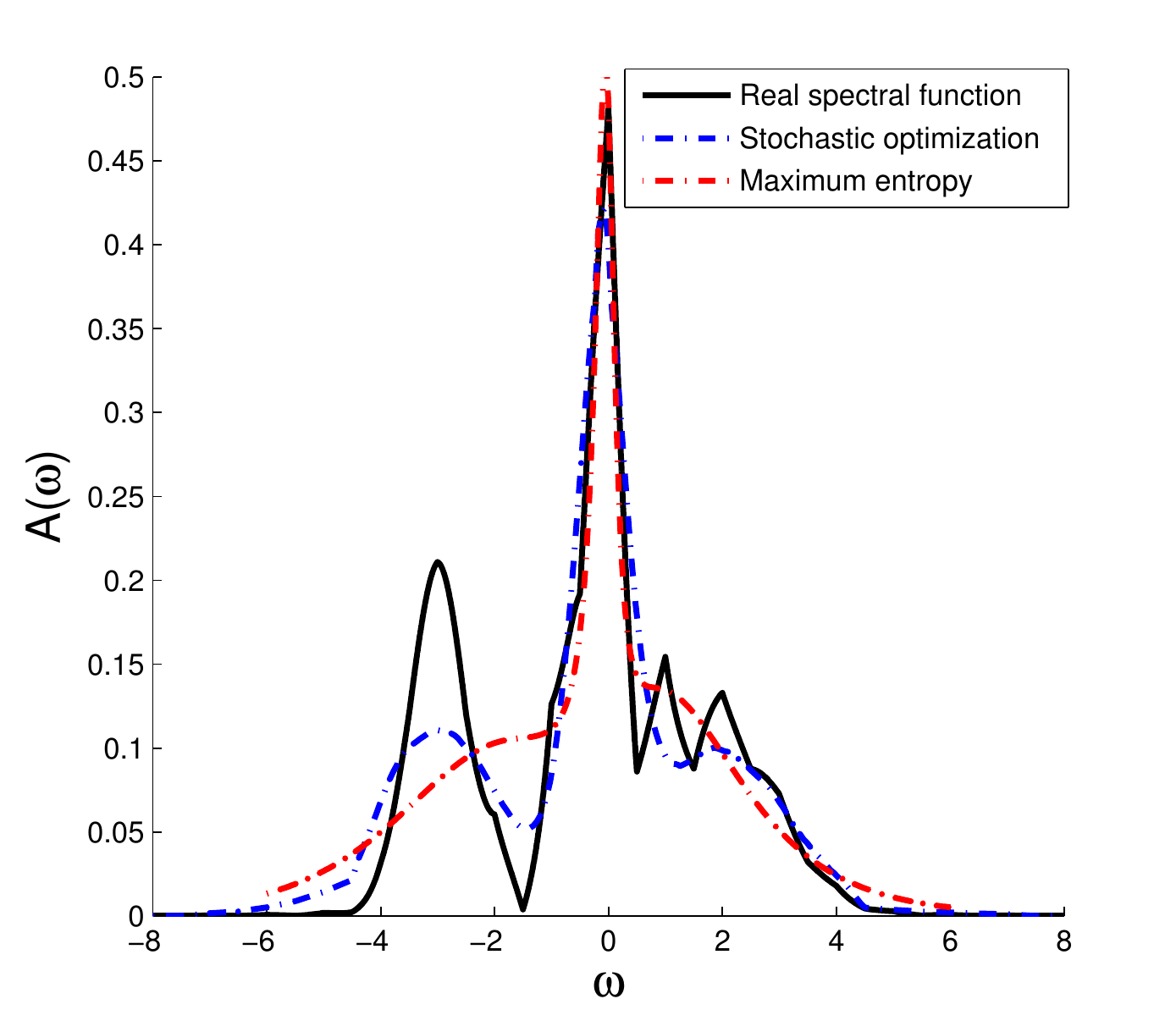}
\end{center}
\caption{Example 1, case 1. Comparison between the true spectrum $\bar{A}
(\omega)$ (black solid curve) with the results of FESOM (blue dash-dotted
line), using a $\chi^2$ threshold $\epsilon=0.05$, and MaxEnt (red dash-dotted
line) for low quality data. Here we have used a noise amplitude of 0.1 to
generate the 1000 samples for the input data. }\label{Ex1_Comparison}
\end{figure}

In order to benchmark our FESOM variant against the state-of-the-art, we compare
in Fig.~\ref{Ex1_Comparison} the results of our approach with correlated noise
and 100 realizations (blue dash-dotted line) with the spectrum obtained from
the MaxEnt procedure described in Sec.~\ref{maxent} (red dash-dotted line).
Here we have used the same 1000 generated samples of the input data
$G(i\omega_n)$ in both FESOM and MaxEnt calculations and a Gaussian default model
for the MaxEnt. For this particular case of low quality data, one sees that the
MaxEnt result only captures the central peak, while the peaks at higher
frequency on either side are washed out. Due to the large $\sigma_{\rm noise}$
of the data, the MaxEnt underfits the data and puts more weight on the entropy
term $S[A]$ in order to minimize the deviation from the Gaussian default model.
In contrast, the FESOM method is able to resolve the higher frequency structure
reasonably well despite the low quality of the data.

\begin{figure}[h!]
\begin{center}
\includegraphics[scale = 0.6]{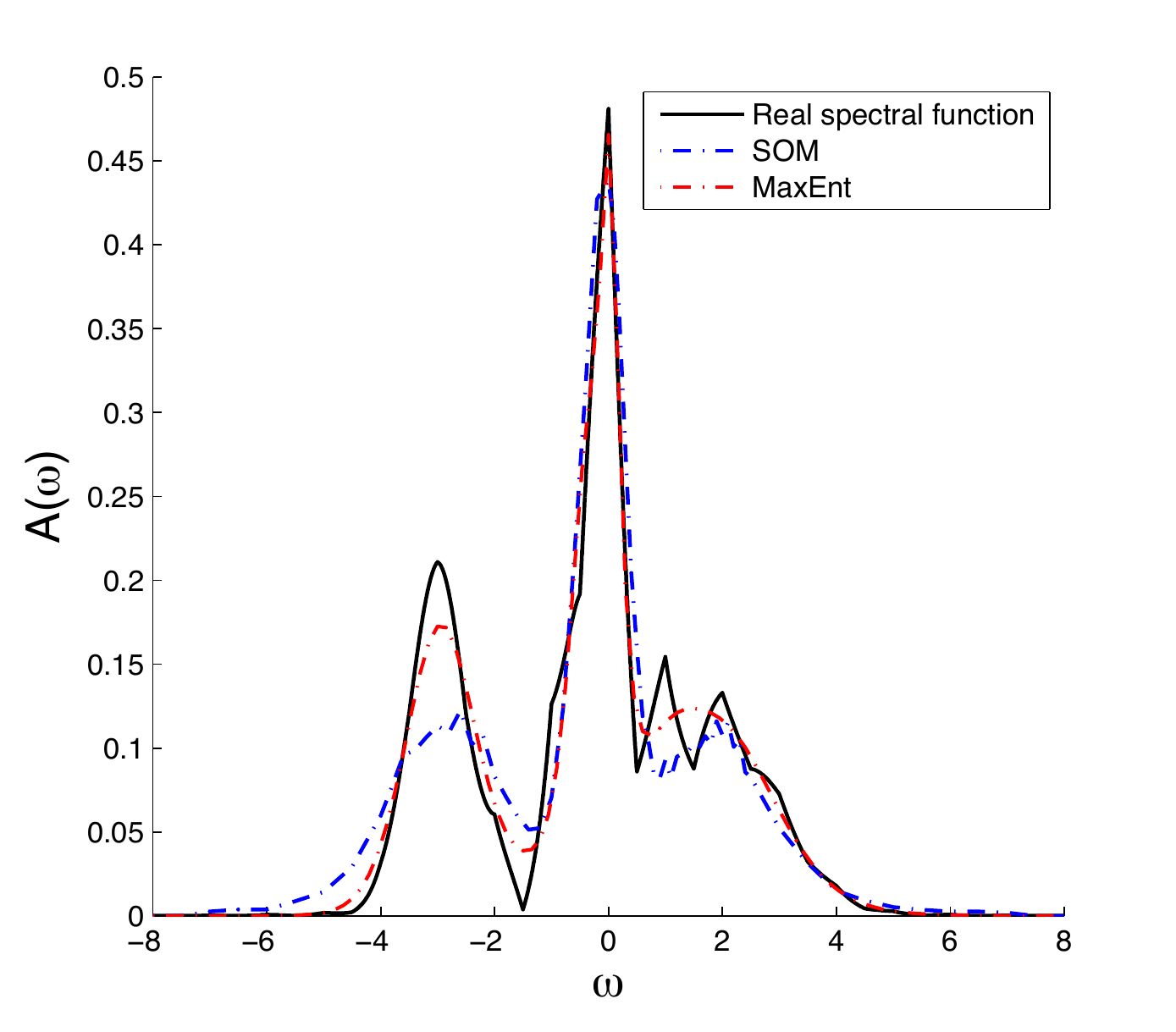}
\end{center}
\caption{Example 1, case 2. Comparison between the true spectrum $\bar{A}
(\omega)$ (black solid curve) with the results of FESOM (blue dash-dotted
line), using a $\chi^2$ threshold $\epsilon=0.05$, and MaxEnt (red dash-dotted
line) for high quality data. Here we have used a noise amplitude of 0.001 to
generate the 1000 samples for the input data. }\label{Ex1_Comparison_case2}
\end{figure}

To study the dependence on data quality in more detail, we have generated a
second set of higher quality input data for the same problem by setting the
noise amplitude $\sigma_{\rm noise}=0.001$. As one sees from
Fig.~\ref{Ex1_Comparison_case2}, here, the MaxEnt gives a much better result
with very good resolution of the structures at higher frequencies. The smaller
$\sigma_{\rm noise}$ forces a better $\chi^2$ fit of the input data and less
similarity with the default model. Fig.~\ref{Ex1_Comparison_case2} also shows
the FESOM result for this case, for which we have used the same $\chi^2$
threshold $\epsilon=0.05$ as in the case of the lower quality data displayed
in Fig.~\ref{Ex1_Comparison}. One sees that the FESOM result is almost
identical to the case with lower quality. \footnote{We have also tried to run
the FESOM optimization with smaller threshold $\epsilon=0.001$, but found that
this leads to prohibitively long runtimes.} We stress that the inferiority of
our FESOM result relative to the MaxEnt in the case of high quality data does
not necessarily reflect a general disadvantage of the traditional SOM
framework, but is likely a result of the simplifications we introduced in our
variant to make the algorithm more efficient. Returning to the case of weak
data in Fig.~\ref{Ex1_Comparison}, we conclude that for cases of low quality
data, for which the MaxEnt procedure tends to underfit the data in the absence
of a good default model, the FESOM approach can provide results that capture
the true spectral function in much more detail.

\begin{figure}[h!]
\begin{center}
\includegraphics[scale = 0.6]{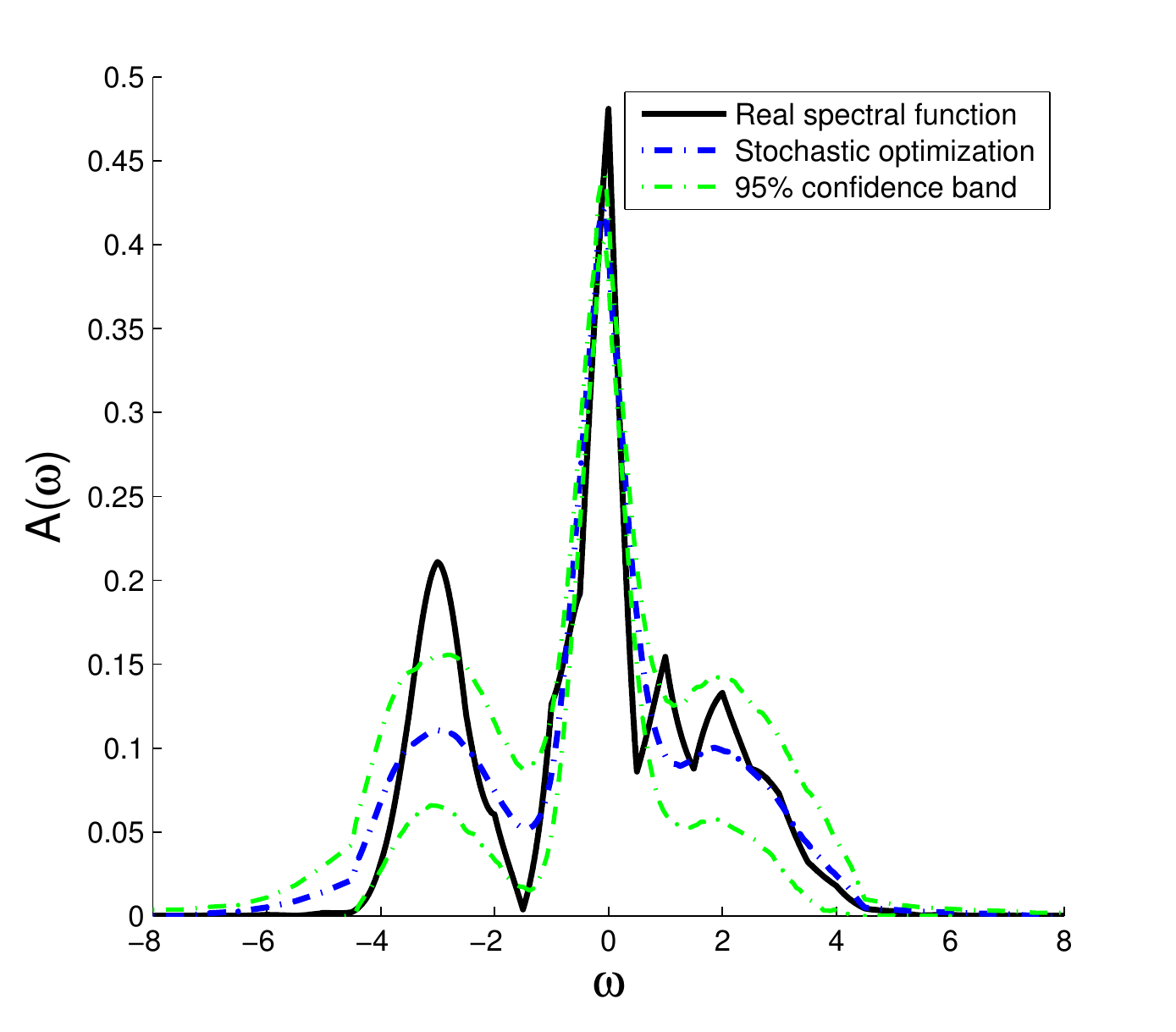}
\end{center}
\caption{Example 1. True spectrum $\bar{A}(\omega)$ (black solid line),
FESOM simulated spectrum $\bar{A}(\omega)$ (blue dash-dotted line) and 95\%
confidence band in FESOM (green dash-dotted line). Here we have used a
$\chi^2$ threshold $\epsilon=0.05$.}\label{Ex1_Confidence}
\end{figure}

As noted, one strength of the stochastic optimization is that one has
information of the confidence interval for all frequencies, while MaxEnt only
allows to determine the uncertainty of the solution integrated over a finite
frequency interval \cite{jarrellBook}. In Fig. \ref{Ex1_Confidence}, we plot
the FESOM simulated spectral function for the low quality input data with it's
$95\%$ confidence region. We can see from the figure that in the frequency
region $[2, 4]$ there are two small peaks in the true spectral function and
neither FESOM and MEM could resolve both peaks well.  However, the large width of
the FESOM confidence band in in this region indicates strong fluctuations in the
different realizations, which in turn could be a signal of possible fine
structure in the true solution.

\subsection*{Example 2.}

\begin{figure*}[ht!]
\begin{center}
\includegraphics[scale = 0.31]{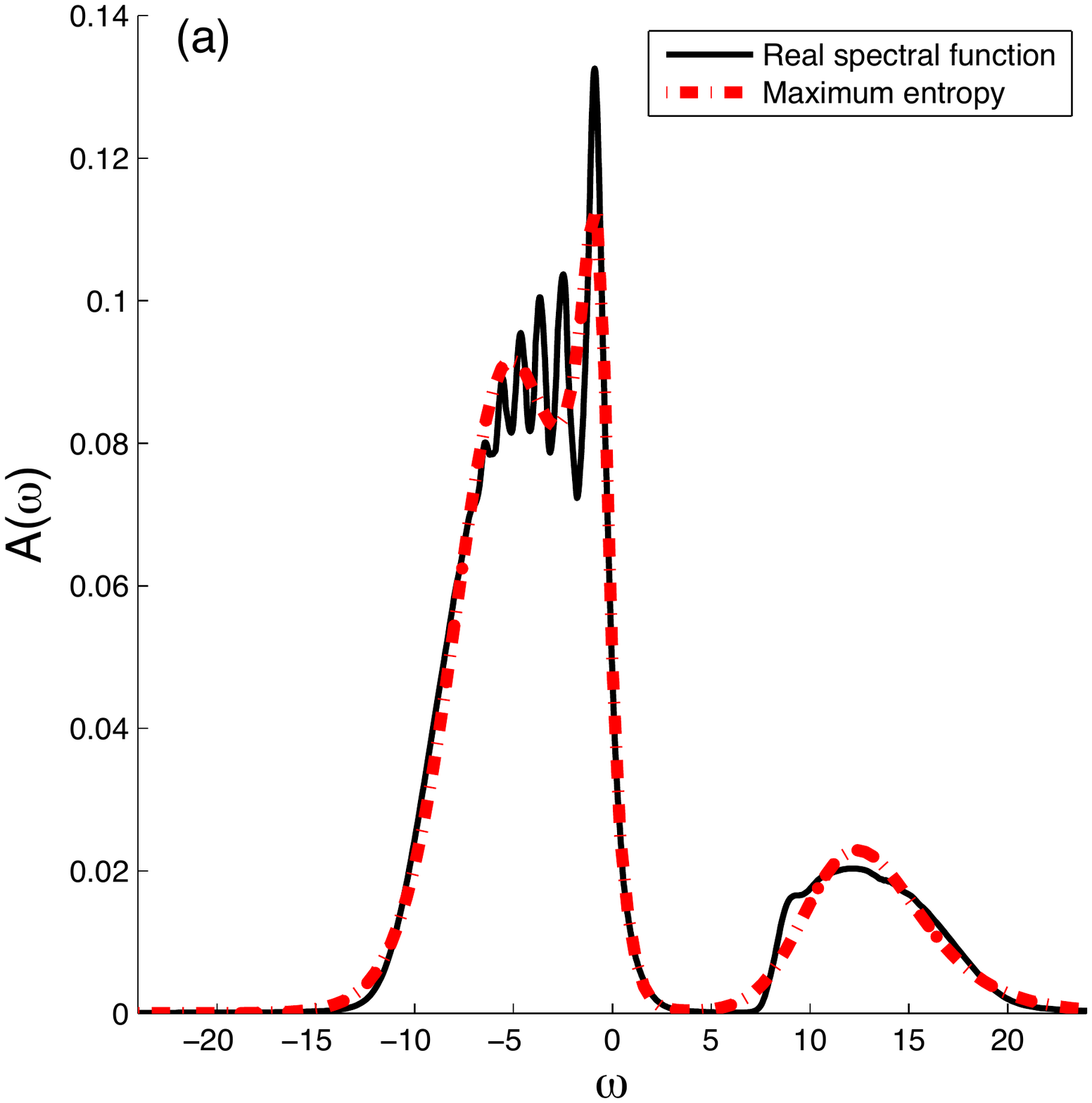}
\includegraphics[scale = 0.31]{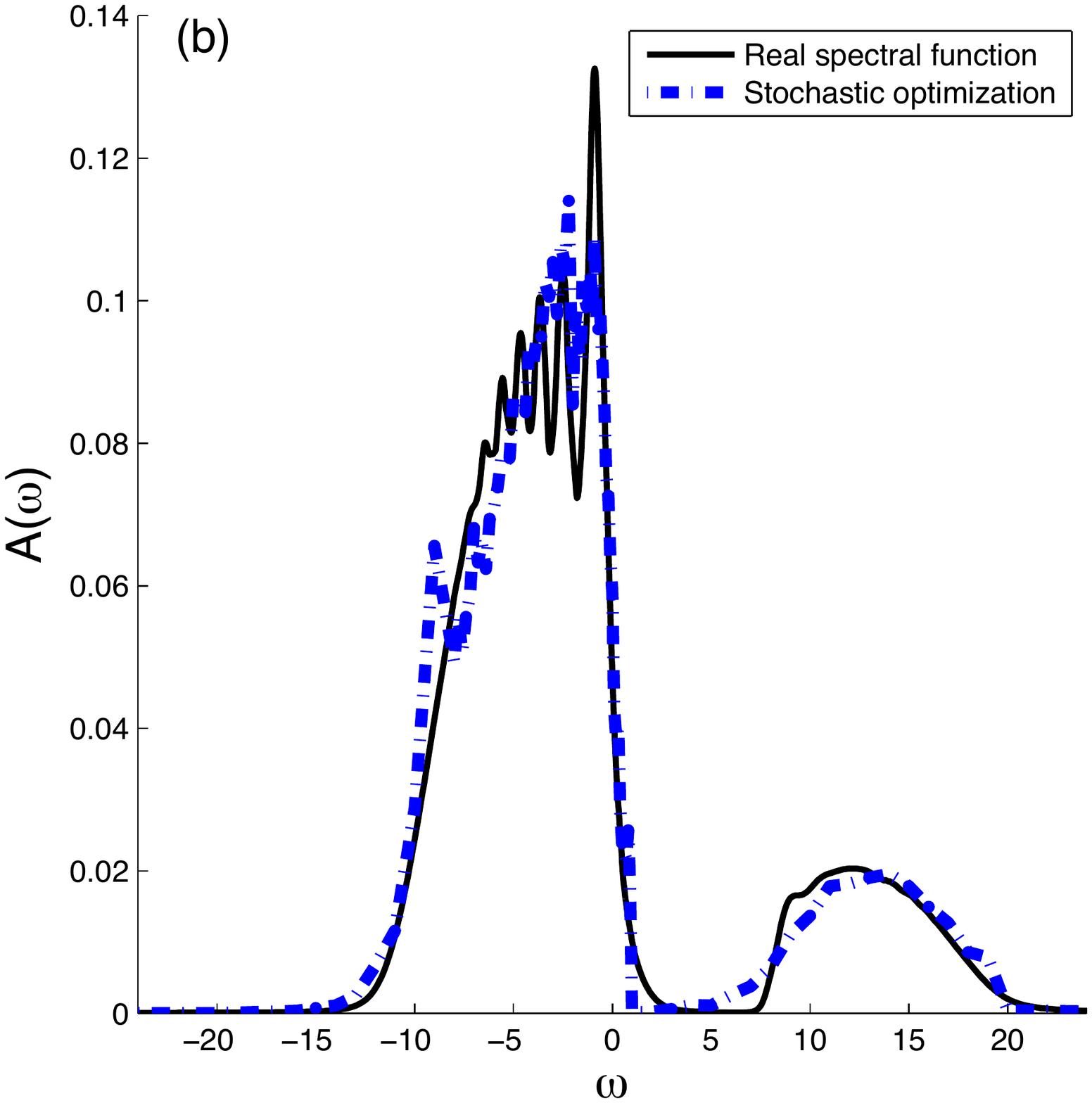}
\includegraphics[scale = 0.31]{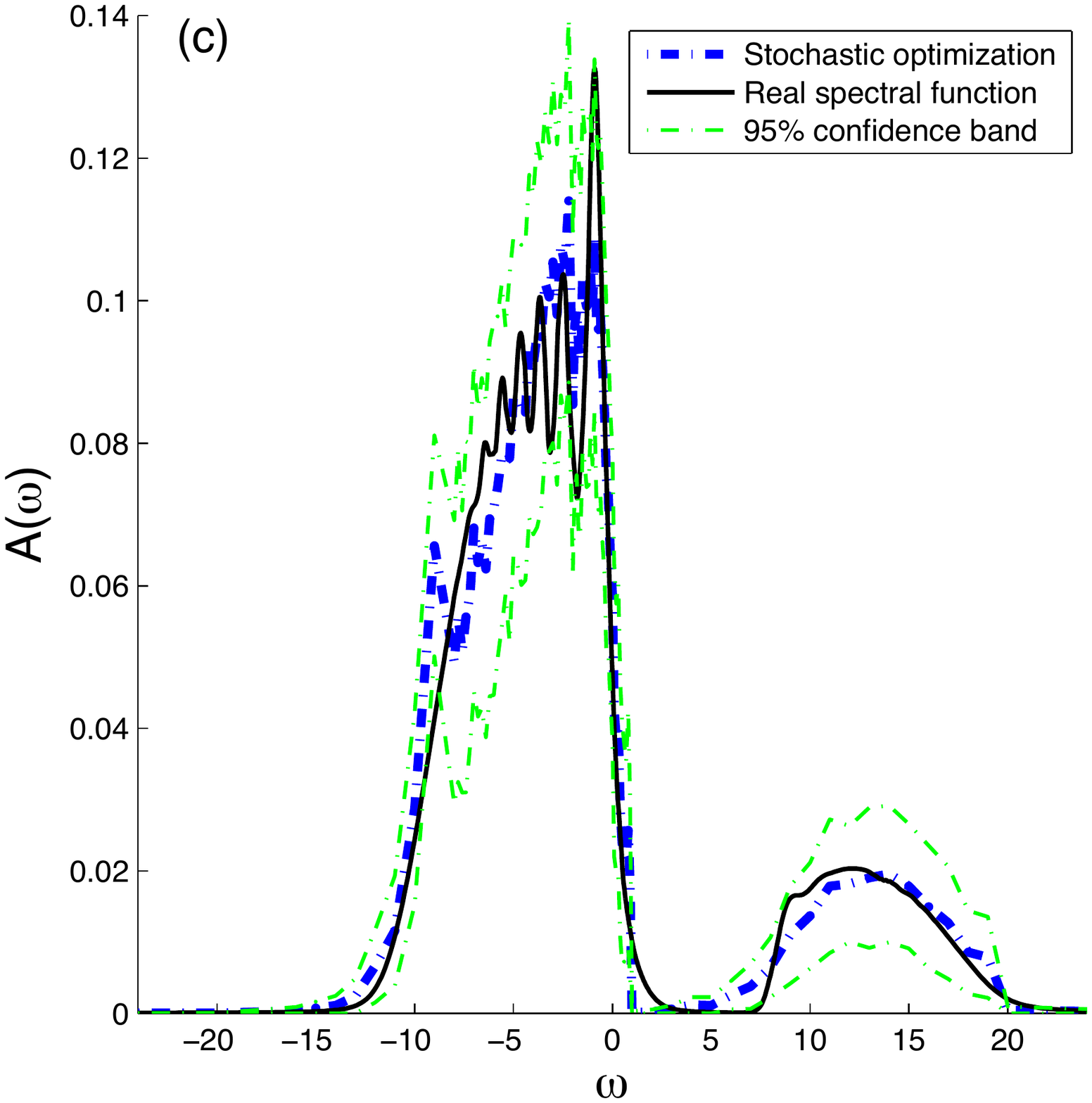}
\end{center}
\caption{ Example 2. Comparisons between true spectrum (black solid line) and
	the results of MaxEnt (a) and FESOM (b), for which we have used a $\chi^2$
	threshold $\epsilon=0.001$. Panel (c) shows the 95\% confidence band
	obtained in the FESOM. The real spectral function is obtained from a
	DMFT/NCA calculation of a 2D Hubbard model in the antiferromagnetic state
	with $U=16t$, $\langle n \rangle = 0.95$ and $T=0.29$.}
\label{Ex2_Comparison_1}. \end{figure*}

We now consider an example for which the true spectral function $A(\omega)$ is
known and generated from a simulation of a microscopic model, and the input
data $G (i\omega_n)$ is again calculated from $A(\omega)$ through Eq.~\eqref
{eq:fredholm}. Specifically, we consider a two-dimensional Hubbard model on a
square lattice with nearest-neighbor hopping $t$ and Coulomb repulsion $U$
described by the Hamiltonian
\begin{align}
\label{eq:Hubbard}
H=-t\sum_{\langle ij\rangle} c^\dagger_{i\sigma}c^{\phantom\dagger}_{j\sigma} +
U\sum_i n_{i\uparrow}n_{i\downarrow}\,.
\end{align}
Here, $c^{\dagger}_{i\sigma}$ creates and $c^{\phantom\dagger}_{i\sigma}$
destroys an electron with spin $\sigma = \uparrow,\downarrow$ on site $i$ and
$n_{i\sigma}=c^\dagger_{i\sigma}c^{\phantom\dagger}_ {i\sigma}$ is the
corresponding number operator. We use the dynamical mean-field theory (DMFT)
\cite{georges96} together with a non-crossing approximation (NCA) \cite{nca} to
obtain the local spectral function $A(\omega)$ in the antiferromagnetic state.
For the majority spin, the local spectral function $A (\omega)$ we obtain is
shown as the black line in Fig.~\ref{Ex2_Comparison_1}. Here we have used
$U=16t$ and set the filling to $\langle n\rangle=0.95$ and temperature
$T=0.29$. One sees the lower and Hubbard bands at negative and positive
frequencies, respectively, as well fine structure with multiple peaks in the
lower Hubbard band. These resonances reflect the bound states of a hole
propagating in an antiferromagnetic background \cite{Str92}.

From this $A(\omega)$, we again generated 1000 samples of the
input data $G (i\omega_n)$ via Eq.~\eqref{eq:fredholm} as in the previous
example by adding noise with standard deviation 0.001. The same samples were
then used in MaxEnt and FESOM to calculate an estimate of $A(\omega)$. The
$\chi^2$ threshold we have used for the FESOM simulation was set to
$\epsilon=0.001$.

Fig.~\ref{Ex2_Comparison_1} compares the MaxEnt result (left panel) and the
FESOM result (middle panel) with the true spectrum. Here one sees that both
approaches capture the lower and upper Hubbard bands equally well. It is
obvious that the MaxEnt has difficulty resolving the fine structure in the
lower Hubbard band at negative frequencies. It captures the first dominant
peak at $\omega=0$, but fails to reproduce the multiple peaks at lower
frequencies. In comparison, the FESOM estimate also has the leading peak, but in
addition shows fluctuations at lower (negative) frequencies, reminiscent to
some extent of the multi-peak stucture in the true $A(\omega)$. These
fluctuations are also seen in the FESOM 95\% confidence band plotted in the
right panel, indicating their presence in a large fraction of the FESOM
realizations. Furthermore, the large width of the confidence band in this
region is a further sign of the fine structure that is present in the true
solution. At higher negative frequencies, however, the FESOM algorithm finds an
artificial peak near $\omega=-9$, while MaxEnt correctly predicts a smooth
result in this region. 

\subsection*{Example 3.}
We now turn to a real data problem, for which the input data $G(i\omega_n)$ is
generated in a QMC simulation and the true spectral function $A(\omega)$ is
not known. We again solve the 2D Hubbard model of Eq.~\eqref{eq:Hubbard}, but
instead of DMFT with NCA impurity solver we now use a dynamic cluster
approximation (DCA) QMC algorithm \cite{hettler98,maier05}, which allows for the
inclusion of non-local correlations in addition to the local correlations
treated in the DMFT. This is accomplished by mapping
the lattice model onto an effective cluster problem embedded in a dynamic
mean-field host that is designed to represent the rest of the system and
determined self-consistently. In order to solve the effective cluster problem,
we use the continuous-time auxiliary-field QMC algorithm by Gull {\it et al.}
\cite{gull08}.

For this example we have set the Coulomb interaction $U=8t$, the site filling
$\langle n\rangle = 0.95$ and the temperature $T=0.08t$ and we have used a
4-site 2$\times$2 cluster for the DCA calculation. From previous calculations
it is known that these parameters give a local spectral function $A(\omega)$
with a pseudogap \cite{jarrell01}, $A(\omega)$ is partially
suppressed at $\omega=0$, reminiscent of the normal state pseudogap phase of
the underdoped cuprate superconductors \cite{norman}.

After the mean-field host is converged, we performed one additional
iteration in which measurements of $G(i\omega_n)$ are performed and
partitioned into 100 bins with a bin size of 100,000 measurements each. For
the MaxEnt procedure, we diagonalized the covariance matrix and rotated the
data and the kernel into the diagonal frame. Moreover, we used the annealing
technique \cite{jarrellBook}, in which the MaxEnt is performed for a set of
decreasing temperatures and the resulting spectrum is used as a default model
for the next lower temperature. The same set of
100 samples of $G(i\omega_n)$ is then used in both the MaxEnt and FESOM to
determine an estimate of the spectral function $A (\omega)$. For the FESOM
analytic continuation, we have set the $\chi^2$ threshoold $\epsilon=0.001$.

\begin{figure*}[ht!]
\begin{center}
\includegraphics[scale = 0.4]{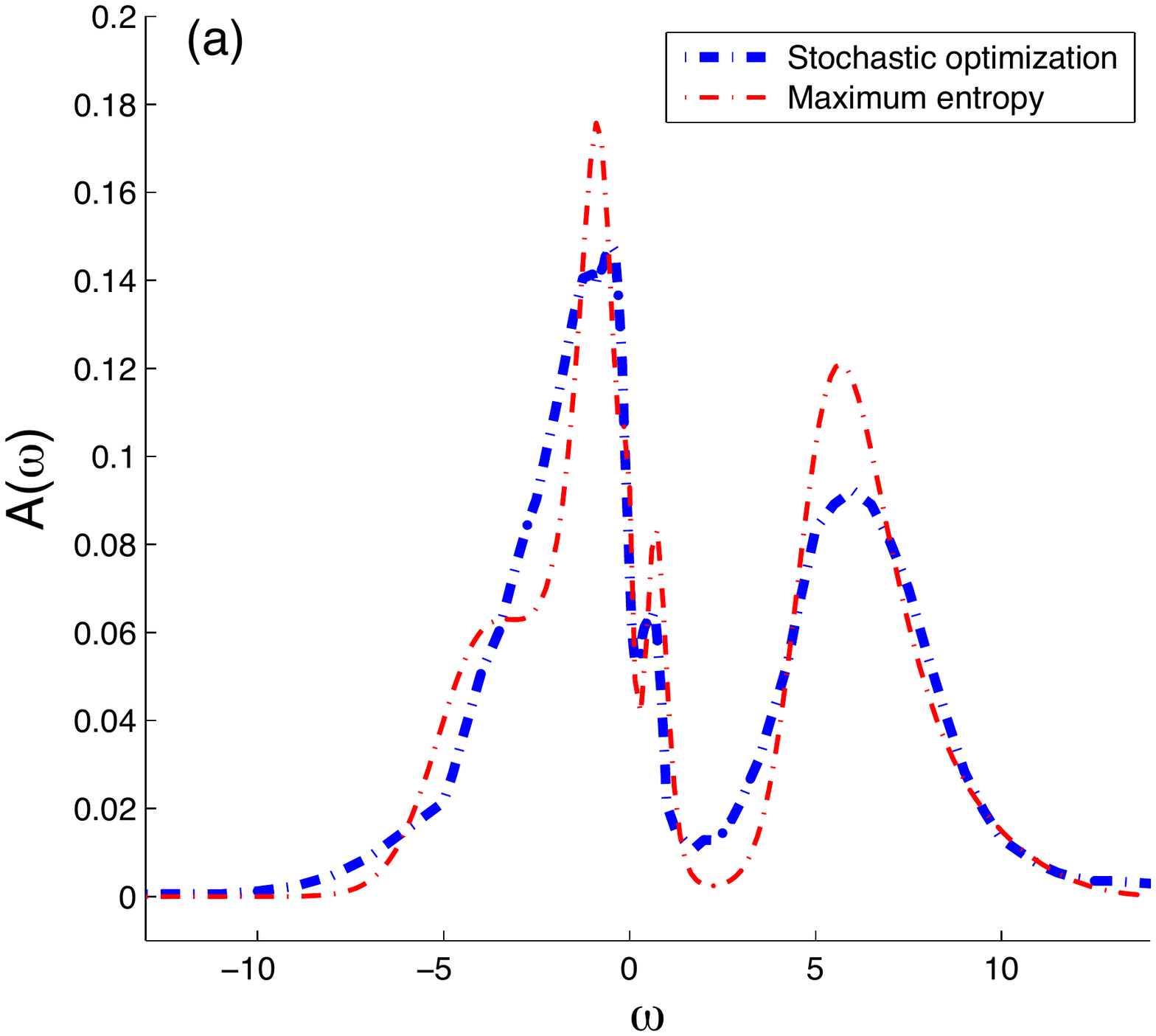}
\includegraphics[scale = 0.4]{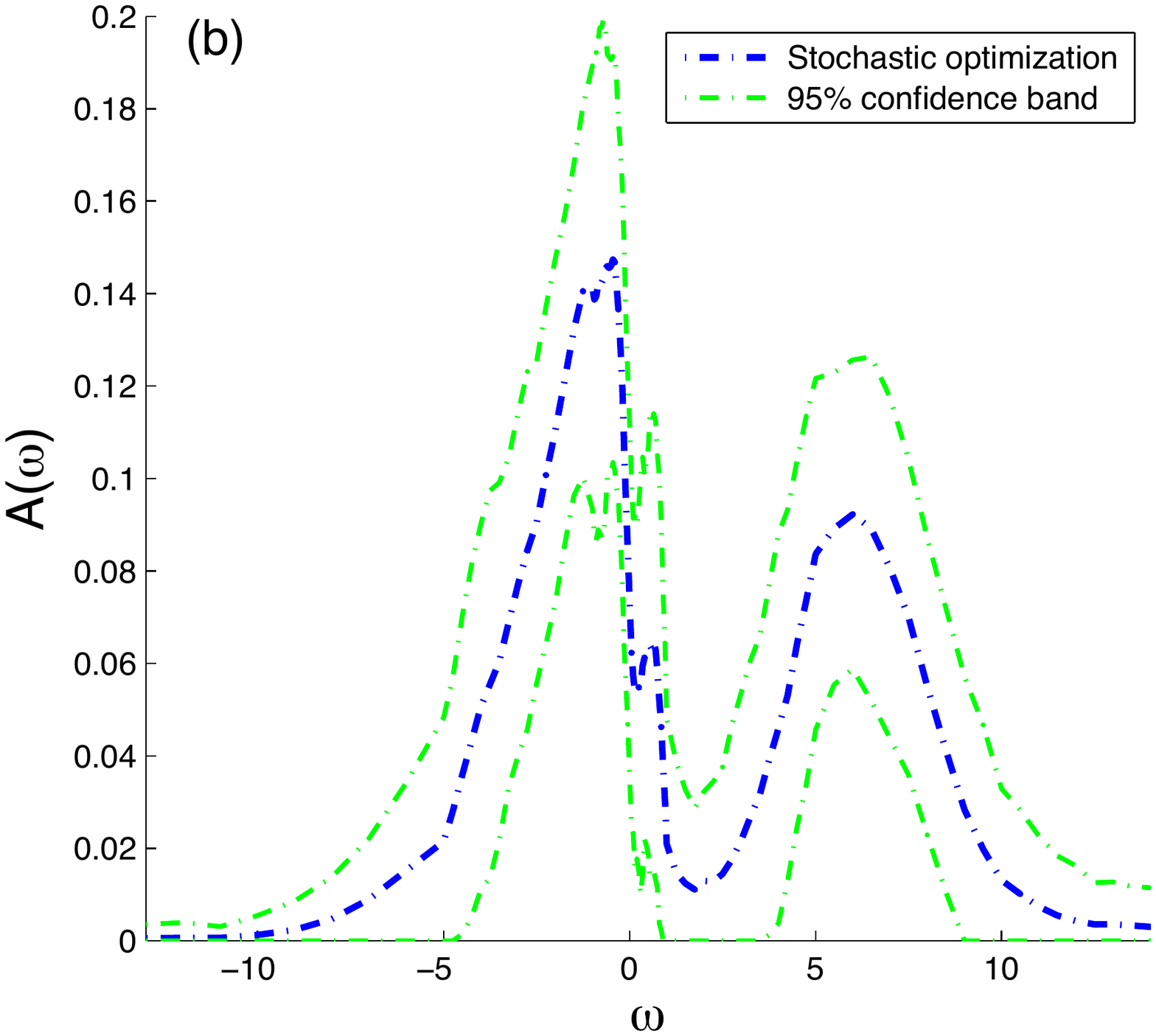}
\end{center}
\caption{Example 3. Spectral function $A(\omega)$ obtained from MaxEnt (a) and
	FESOM (b) (using a $\chi^2$ threshold of $\epsilon=0.001$) analytic
	continuation of DCA QMC data for a 2D Hubbard model with $U=8t$,
	$\langle n\rangle = 0.95$ and $T=0.08t$. }\label{Ex3_Comparison}
\end{figure*}

In the left panel of Fig.~\ref{Ex3_Comparison}, we compare the simulated
spectral function $A(\omega)$ obtained from the MaxEnt (red solid line) with
that of the FESOM calculation (blue dashed line). One again sees the two Hubbard
bands centered below and above $\omega=0$ and split by $\sim U=8t$. For this
case, the MaxEnt result clearly shows more structure in the lower Hubbard
band. Both the MaxEnt and the FESOM resolve the pseudogap feature, manifested as
the dip in $A(\omega)$ at $\omega=0$. But it is much better developed in the
MaxEnt than in the FESOM $A(\omega)$. In addition, the MaxEnt result displays a
shoulder at $\omega\sim -4$, which is not present in the FESOM result.

The right panel of Fig.~\ref{Ex3_Comparison} displays the 95\% confidence
band obtained from the FESOM simulation. The band is unusually wide
even at small $|\omega|$, indicating large fluctuations in the different
realizations. However, once again one sees that the confidence band follows
the same trend as the mean spectrum $\bar{A}(\omega)$. This shows that the
pseudogap feature is present in a large fraction of the FESOM realizations and
therefore likely a feature of the true spectral function.

We also tried the annealing technique for the FESOM simulation. We did find
faster convergence of the optimization procedure in the last step of the
annealing procedure. However, there was no change in the resulting spectrum.
Considering that a separate optimization has to be carried out for each
temperature, the annealing method is less efficient than just running a single
optimization at the lowest temperature and, in contrast to the MaxEnt, does
not provide any improvement in the solution.

% In this example, we present the performance of SOM in solving a real data
% problem. {\color{blue} Please add the physical background of this problem.}

% In Fig. \ref{Ex3_Comparison}, we compare the simulated spectral function
% obtained by using the classic MEM and the SOM. The MEM we use here is
% obtained by using annealing data which uses the spectral function at high
% temperature as the default model for the next lower temperature. The black
% curve is the true synthetic spectral function, the red dashed curve is the
% simulated spectral function obtained by using MEM and the blue dashed curve
% is the simulated spectral function obtained by using the SOM. We can see
% from this figure that both methods capture the shape of the spectral
% function including the pseudo dip around $0$ frequency. {\color{blue} Please
% explain why this is importnat. } In Fig. \ref{Ex3_Confidence} we also plot
% the $95\%$ confidence band of the SOM. We can see that there's a clear dip
% in the confidence band which further confirming this feature in the spectral
% function.

% It is worth to point out although both methods give similar results, the FESOM does not depend on the initial condition (default model) while the MEM needs several annealing steps to reach this result. Therefore, the SOM is advantageous to MEM when the annealing data is hard to obtain.

\section{Summary and conclusions}
To summarize, we have introduced, analyzed and benchmarked against Maximum
Entropy a fast and efficient variant of the stochastic optimization method
introduced by Mishchenko {\it et al.} \cite{Mishchenko00}, called FESOM, for
the analytical continuation of imaginary frequency QMC data $G(i\omega_n)$, an
ill-posed problem that remains a challenging barrier in connecting theory with
experiment. In contrast to the routinely used Maximum Entropy method, which
introduces a default model to regularize the problem, the stochastic
optimization method only uses minimal prior information for the quantity of
interest, the spectral function $A(\omega)$, and does not introduce a default
model. The basic idea of the SOM approach is to use several parallel
optimization procedures to otain a large set of equally likely estimates of
the spectrum and determine the final spectral function $A(\omega)$ as an
average over these samples. The optimization procedure minimizes the $\chi^2$
misfit between the QMC data and the modeled data by sequentially and randomly
proposing global changes to a test spectral function. A combination of three
characteristics of our FESOM implementation results in a more efficient and less
complex algorithm than the previous implementation: (1) It uses a fixed
frequency grid just like MaxEnt. (2) It only allows for proposal updates that
lower $\chi^2$ and does not permit temporary increases. (3) It uses a Gaussian
process to update the spectral function, in which the noise added to adjacent
frequencies is correlated. Characteristic (3) can be viewed as an implicit
regularization that results in much smoother individual estimates and
therefore a smaller number of realizations required to get a smooth average.

We have applied this algorithm to three representative test case problems and
compared the results with those obtained from MaxEnt. For two of these
problems, the true spectral function $A(\omega)$ was known and used to generate
a noisy set of input data $G(i\omega_n)$. For the third case, we  used QMC data
for $G(i\omega_n)$ obtained from DCA simulations of a 2D Hubbard model.  For
these problems, our FESOM algorithm generally gave similar spectra to those
obtained from MaxEnt. For good quality data with weak noise, we found that the
MaxEnt procedure gives much better results than the FESOM method, while for poor
quality data the situation is reversed. In this case, the MaxEnt tends to
underfit the data, while the FESOM procedure gives a much better result.
Generally, we found that in contrast to MaxEnt, the quality of the FESOM depends
very little on the quality of the input data. In addition, the FESOM provides
information on the confidence of the resulting spectral function $A(\omega)$
for each frequency $\omega$, in contrast to MaxEnt, which only gives this
information for a finite interval in frequency.

For the test case problems we have studied, the stochastic optimization
required on average about 1-2 minutes on a single core (2.2 GHz Intel Core i7)
to optimize a single realization and a total runtime of $\sim$ 2 core
hours to produce the final spectrum as the average of 100 realizations. While
the total runtime is about an order of magnitude longer than that of the MaxEnt
procedure with annealing, per realization it is roughly of the same order.
Trivial parallelization of the stochastic optimization over different
realizations will therefore result in similar runtimes. We therefore believe
that our implementation of the stochastic optimization technique provides a
viable alternative to the MaxEnt procedure for the analytic continuation of QMC
data, especially for cases with poor data quality.

\section*{Acknowledgements}
This research was sponsored by the Laboratory Directed Research and Development
Program of Oak Ridge National Laboratory, managed by UT-Battelle, LLC, for the
U. S. Department of Energy. V.W.S. acknowledges support from AFOSR (grant
FA9550-15-1-0445) and ARO (grant W911NF-16-1-0182).

\end{document}